\documentclass[conference]{IEEEtran}
\IEEEoverridecommandlockouts

\usepackage[utf8]{inputenc}
\usepackage[T1]{fontenc}
\usepackage{cite}
\usepackage{url}
\usepackage[utf8]{inputenc}
\usepackage[T1]{fontenc}
\usepackage{subcaption}
\usepackage[stable]{footmisc}
\usepackage{paralist}
\usepackage{caption}
\usepackage{graphicx}
\begin{document}

\title{V-SHiNE: A Virtual Smart Home Framework for Explainability Evaluation
\thanks{© 2026 IEEE. This is the author’s accepted manuscript of a paper accepted at the
2026 IEEE International Conference on Pervasive Computing (PerCom 2026).
Personal use is permitted; permission from IEEE is required for all other uses.}}

\author{
\IEEEauthorblockN{Mersedeh Sadeghi}
\IEEEauthorblockA{
\textit{University of Cologne} \\
Cologne, Germany \\
mersedeh.sadeghi@uni-koeln.de \\
}
\and
\IEEEauthorblockN{Simon Scholz}
\IEEEauthorblockA{
\textit{University of Cologne} \\
Cologne, Germany \\
sschol24@smail.uni-koeln.de
}
\and
\IEEEauthorblockN{Max Unterbusch}
\IEEEauthorblockA{
\textit{University of Duisburg-Essen} \\
Essen, Germany \\
max.unterbusch@uni-due.de \\
}
\and
\IEEEauthorblockN{Andreas Vogelsang}
\IEEEauthorblockA{
\textit{University of Duisburg-Essen} \\
Essen, Germany \\
andreas.vogelsang@uni-due.de \\
}
}
\maketitle

\begin{abstract}
Explanations are essential for helping users interpret and trust autonomous smart-home decisions, yet evaluating their quality and impact remains methodologically difficult in this domain. V-SHiNE addresses this gap: a browser-based smart-home simulation framework for scalable and realistic assessment of explanations. It allows researchers to configure environments, simulate behaviors, and plug in custom explanation engines, with flexible delivery modes and rich interaction logging. A study with 159 participants demonstrates its feasibility. V-SHiNE provides a lightweight, reproducible platform for advancing user-centered evaluation of explainable intelligent systems.
\end{abstract}


% ----------------------------
\section{Introduction \& Motivation}

Pervasive computing continues to transform daily life. Individuals are increasingly surrounded by embedded and cyber-physical systems (CPS) coupled with digital infrastructure and ubiquitous connectivity~\cite{baheti2011cyber}. These systems enable autonomous, context-aware services that adapt dynamically to users’ needs~\cite{vermasan2014internet}. Emerging smart homes, as one of the prominent applications of modern pervasive systems ~\cite{houze2022generic}, illustrate both their benefits and their challenges: While they offer convenience and efficiency, their autonomous decision-making, complex rules, and opaque machine learning models make them difficult to understand, configure, and trust~\cite{shin2021effects}. 

Given the growing complexity and autonomy of smart homes, developing \textit{explanation} methods and intelligent systems that can \textit{explain} themselves has become increasingly important~\cite{sadeghi2021cases,jakobi2018evolving}. 
Although considerable progress has been made in designing these methods, assessing their usefulness in realistic settings continues to be challenging.

Fully instrumented smart homes are rarely available, labs have only a few devices and require tedious reconfiguration, surveys are scalable but offer little ecological realism. Despite advances in explainability research~\cite{sadeghi2021cases,carneiro2020explainable,jha2022overview}, system metrics such as accuracy or fidelity~\cite{abdul2018trends,miller2019explanation,nunes2017systematic} provide little insight into whether explanations are clear, helpful, or trustworthy. User studies remain essential~\cite{miller2019explanation,doshi2017towards}, yet current methods are costly, intrusive, or ecologically limited. 

In sum, while user-centered evaluation of explanations is recognized as essential, scalable, reproducible, and realistic methods remain lacking. To address this, we present \textbf{V-SHiNE} (Virtual Smart Homes with Intelligent and Explainable Features), a browser-based, first-person smart home simulation framework supporting configurable environments and flexible integration of explanation.

\begin{figure}
    \centering
    \captionsetup{font=small}\includegraphics[width=0.90\linewidth]{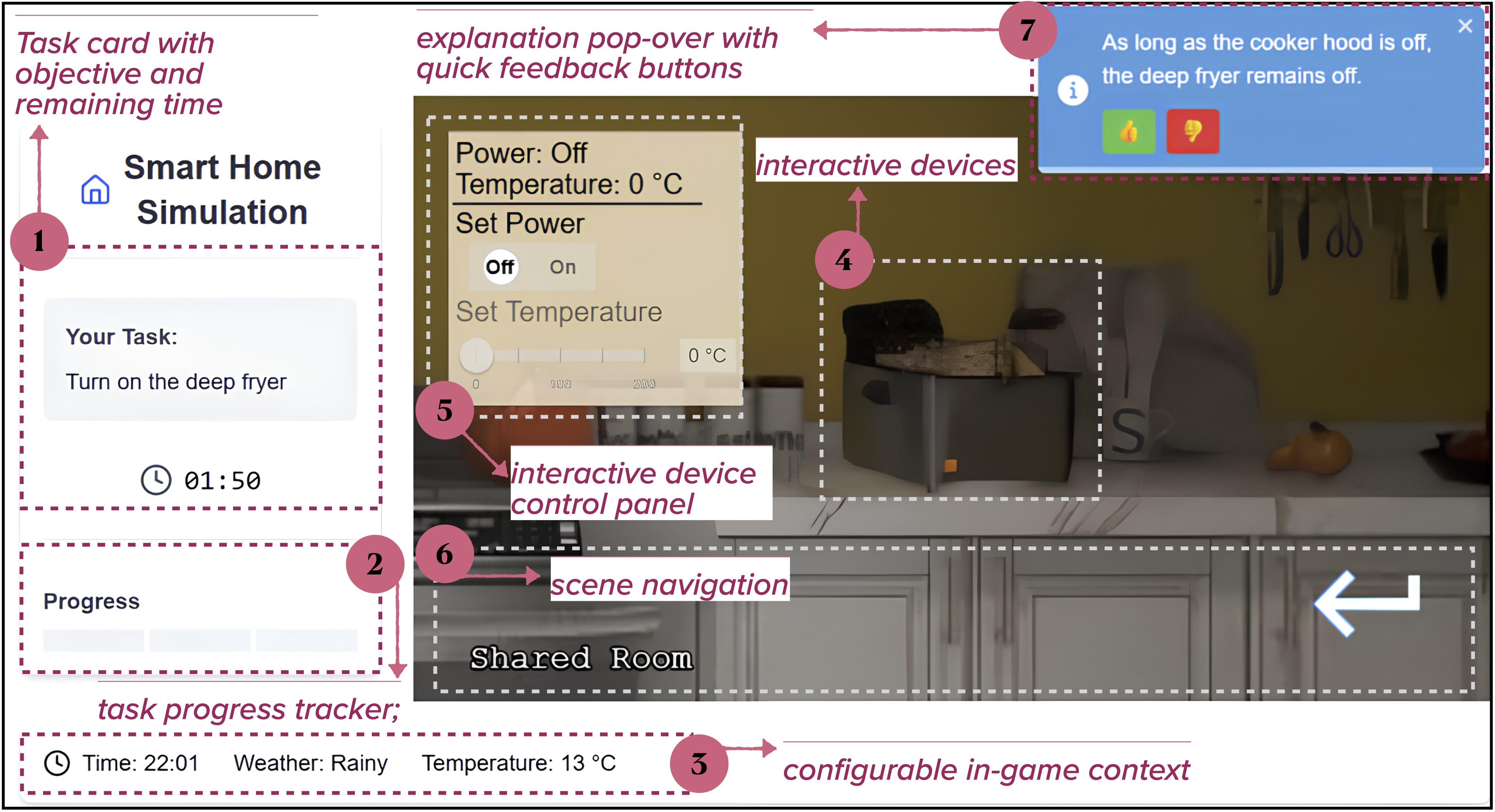}
    \caption{V-SHiNE interface layout.}
    \label{fig:screenshot}
\end{figure}

With V-SHiNE, researchers can quickly design interactive smart home scenarios. Using JSON files, they define apartments, devices, and automation rules, including conflicting behaviors. They can also implement explanations for participants to view during the simulation, either by defining them directly within these JSON files or by connecting an external explanation engine via an API that dynamically fetches them based on various contexts. Participants access V-SHiNE through their browser. As shown in Figure~\ref{fig:screenshot}, they navigate from a first-person perspective, interact with devices, and receive explanations in different formats (push, pull, interactive). V-SHiNE logs all interactions and collects ratings, giving researchers both subjective and behavioral data.

Our demonstration replicates this workflow: configuring an environment, triggering system behaviors, observing explanations, and analyzing feedback. The goal is to show how V-SHiNE turns a once costly and constrained evaluation process into a lightweight, reproducible, and scalable workflow for explainability research.

\section{The V-SHiNE Framework}
\begin{figure}[b]
  \centering
  \includegraphics[width=.8\linewidth]{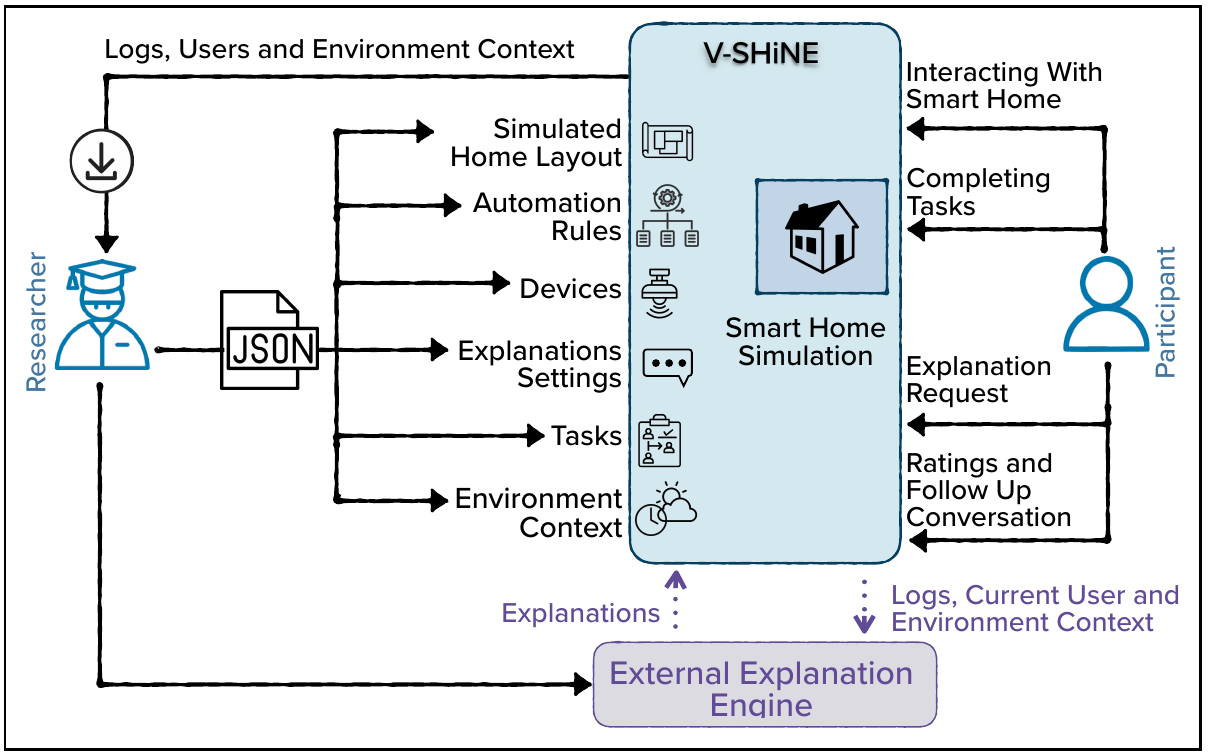}
  \caption{High-level overview of the V-SHiNE framework.}
  \label{fig:overall}
\end{figure}

%\subsection{Overview}
V-SHiNE is implemented as a distributed web application designed for interactive smart home simulation studies. At its core, it provides a lightweight testbed for configuring realistic environments, simulating device behaviors, and integrating explanation engines. The architecture is modular, consisting of a \emph{frontend}, \emph{backend}, a persistent \emph{database}, and optional connections to \emph{external explanation engines}. These components communicate through real-time WebSocket events (via Socket.IO) and standard REST APIs, enabling flexible integration and interactive responsiveness. Figure~\ref{fig:overall} illustrates the high-level overview of the V-SHiNE framework.

\subsection{Key Features}
\textbf{A Highly Configurable Virtual Testbed.}  
The \emph{frontend} combines React components for study orchestration with the Phaser game engine for rendering interactive environments. Researchers configure scenarios using JSON files that define room layouts, devices, and automation rules. Devices become interactive objects in the simulation, and user actions are sent as real-time events to the backend, where rules are evaluated. %This setup allows easy reconfiguration without physical lab changes.

Beyond this, V-SHiNE supports a detailed \emph{game schema} that lets experimenters control nearly every aspect of the environment. The JSON specification defines:

\begin{compactitem}
  \item \textbf{Room geometry and layout}: specify room boundaries, connections (doors), and object placement so that the virtual environment mirrors intended floorplans.  
  \item \textbf{Device types and attributes}: includes specifying devices with their types (e.g., heater, light), initial states, and allowed user interactions (e.g., on/off, opening/closing windows), along with contextual constraints like temperature thresholds or device dependencies
  \item \textbf{Virtual control panels}: a per-device control panel (see block No. 5 in Fig.~\ref{fig:screenshot}) dynamically rendered from an abstract interaction model defined in the JSON configuration. To generalize across device classes, we follow the W3C Thing Description (TD)~\cite{w3c-td} and its TDeX-based extension to map interaction types to GUI elements~\cite{baresi2018tdex}.  
  \item \textbf{Rules and automations}: express condition-action rules that simulate automation behavior (e.g., ``if window is open and outside temperature <15$^\circ$C, then heater must stay on''). These rules are evaluated by the backend whenever relevant state changes occur.  
  \item \textbf{Triggers and events}: embed triggers that cause state changes or exceptions (e.g., device failures, environment transitions) at set times or in response to user actions.  
  %\item \textbf{Explanation bindings}: assign explanation templates or API endpoints to rule violations or unexpected behaviors, and specify the delivery types (push, pull, interactive).  
  \item \textbf{Task definitions}: encode experiment tasks (e.g., ``turn off the heater after 3 minutes'' or ``investigate why a device did not respond'') to guide participant interaction flows.  

  \item \textbf{Contextual variables and session personalization}: define environment-specific variables (e.g., weather, temperature, occupancy) and user-level context (e.g., novice vs. expert, treatment group) through base64-encoded session parameters. These variables can dynamically influence rule evaluation, device behavior, and explanation delivery.
\end{compactitem}
%When the frontend loads the JSON schema, it renders the room layout and initializes all devices with their starting states. As users act, events propagate to the backend, where rule logic enforces consistency and triggers explanations when needed. Since the schema is fully declarative, experimenters can easily create new environments by editing JSON rather than modifying core system code.

\textbf{A Modular and Flexible Explanation System.}  
The back-end, built with Next.js and Socket.IO, manages session state, device interactions, and communication with explanation engines. Explanations can be delivered directly from V-SHiNE or integrated from external engines through REST or WebSocket interfaces. Researchers can explore different delivery models (push, pull, interactive) by adjusting configuration files (explanation schema) rather than modifying the core system.

\textbf{Comprehensive Data Logging and Feedback.}  
A MongoDB database stores session data, device states, explanation content, ratings, and detailed logs of all user interactions. The system captures both explicit feedback (e.g., ratings) and implicit behavior (e.g., device interactions, task completion times). This enables correlation between subjective perceptions and objective measures, supporting rigorous post-hoc analysis.

\subsection{Testing and Reliability Assurance}

To ensure the robustness and correctness of V-SHiNE’s core components, we maintain a comprehensive Vitest-based (Node.js) test suite covering three categories. Firstly, \textbf{Unit Tests: API routes.} that validates Next.js HTTP endpoints (e.g., session creation, state retrieval, study completion) in isolation. Mocked MongoDB operations and request/response scaffolding ensure correctness without reliance on real infrastructure. Secondly, \textbf{Integration Tests: Socket workflows.} that exercises real-time interactions across modules: session validation, device updates, rule evaluation, broadcasting, explanation requests, and data persistence. They simulate full event chains through Socket.IO, covering multi-service workflows end-to-end. Finally, \textbf{Configuration Validation Tests.} These tests verify the integrity of configuration files (e.g., JSON schemas for game environment definitions and explanation templates), ensuring that user-defined scenarios comply with expected formats before runtime.  

Overall, the test suite comprises 21 test files with roughly 500 test cases, targeting a minimum coverage threshold of 70\% across branches, functions, lines, and statements. The testing harness employs mocks for MongoDB, a Socket.IO test infrastructure, and standardized test fixtures (sessions, device states, API payloads).

%Overall, the test suite comprises 21 test files (\(\sim\)500 cases) with a \(\ge\)70\% coverage target (branches, functions, lines, statements) and uses MongoDB mocks, a Socket.IO test rig, and standardized fixtures (sessions, device states, API payloads). Notably, while the backend logic, API, and socket layers are covered, the visual frontend currently lacks automated testing due to its interactive nature. An end-to-end testing (e.g., via browser automation tools) is a potential future enhancement. 

%The test suite comprises 21 files (\(\sim\)500 cases) with a \(\ge\)70\%  coverage target, using MongoDB mocks, a Socket.IO test rig, and standardized fixtures. While the backend logic, API, and socket layers are covered, the visual frontend currently lacks automated testing due to its interactive nature, and end-to-end testing is a potential future enhancement.

Researchers relying on V-SHiNE can thus have confidence in the consistency of core behaviors of their study thanks to built-in testing coverage.
% ----------------------------
\section{Demo Walkthrough}
During the demonstration, attendees will:  
\textbf{(1)} Open a preconfigured smart-home scenario in the browser. \textbf{(2)} Interact with devices and experience both expected and unexpected system behaviors. \textbf{(3)} Receive explanations through different delivery models (automatic, on-demand, interactive). \textbf{(4)} Provide feedback using built-in rating mechanisms. \textbf{(5)} Observe how all interactions and system events are logged. %.  

\textbf{Load Smart Home Scenario.}  
The demo begins by loading a preconfigured environment defined in a JSON file. This file specifies the layout of rooms, the devices and automation rules. Once loaded, the environment renders in a browser, presenting participants with a first-person view of the smart home. This setup can be easily adapted to new studies by modifying the JSON configuration rather than restructuring a physical lab.

\textbf{Interact with devices.}  
Participants can move through the virtual home and interact with devices such as heaters, thermostats, or lights. Actions may succeed or fail depending on the environment state, for example, attempting to switch off a heater may be blocked if the indoor temperature is below a threshold. These interactive tasks create natural opportunities for explanations to be triggered, mimicking realistic smart home experiences.

\textbf{Receive explanations.}  
When confronted with unexpected outcomes, explanations in V-SHiNE can be delivered in three modes: \textit{pull} (shown upon user request), \textit{push} (shown automatically), and \textit{interactive} (adds a chat input for user queries). For instance, when unable to turn off the heater, the system may explain: \emph{``The indoor temperature is lower than 15°C.''} A follow-up query could reveal a deeper cause: \emph{``The window is open and the outside temperature is below 15°C.''}% Such interaction illustrates how V-SHiNE supports different explanation strategies and enables evaluation of their clarity and usefulness in realistic contexts.

\textbf{Provide feedback.}  
Participants can directly rate the explanations through built-in mechanisms (see the thumb up/down symbols in block No. 7 in Figure~\ref{fig:screenshot}). Therefore, subjective impressions are also captured alongside behavioral data.

\textbf{Analyze logs.}  
All actions, requests, and ratings are logged automatically. Researchers can analyze these logs to compare explanation strategies, study user behavior, and correlate subjective and objective measures of effectiveness. An exhastive list of events logged by V-SHiNE can be found in the documentation (\label{note:doc}\url{https://exmartlab.github.io/SHiNE-Framework/}). 
%Figure~\ref{fig:analysis} shows an example of a comparison between three types of explanation with respect to the number of user interactions and the number of completed tasks the participants achieved in the experiment. This analysis is based on data logged and exportet by V-SHiNE.

%\begin{figure}
%    \centering
%    \begin{subfigure}{0.5\columnwidth}
%      \includegraphics[width=\linewidth]{Figures/log_count.png}
%        \caption{Interaction Steps}
%    \end{subfigure}%
%    \begin{subfigure}{0.5\columnwidth}
%    \includegraphics[width=\linewidth]{Figures/success.png}
%       \caption{Completed Tasks}
%    \end{subfigure}
%   \caption{Analysis of exported log data}
%    \label{fig:analysis}
%\end{figure}

% ----------------------------
\section{Validation \& Early Experience}
To validate V-SHiNE, we applied it in two distinct contexts.  
First, we conducted a large-scale empirical study with 159 participants evaluating the effectiveness, usability, and impact of context-aware and static explanation strategies generated by our system. A comprehensive report of this study is currently under review.

Second, our documentation includes two illustrative scenarios. The Default Scenario models three everyday smart-home tasks, each governed by a different automation pattern, enabling controlled assessment of explanations for conflicts, constraints, and delayed effects. To ensure neutrality and highlight general applicability, we additionally selected a scenario from an external study with no connection to our research group. Specifically, we re-implemented an example from the CIRCE study~\cite{Reyd2024}, which had validated its causal explanation method solely through system-level simulation. Integrating CIRCE into V-SHiNE illustrates how explanation approaches that lack user-centered evaluation can be seamlessly extended into realistic, task-based user experiments.
%Second, we reimplemented an interactive scenario from a published study by another research group\cite{Reyd2024}, demonstrating how V-SHiNE can generalize across approaches and fill evaluation gaps by enabling realistic, task-based user studies. These results illustrate both the framework's feasibility and its potential for broader adoption.

% ----------------------------
%\section{Related Work}
%Prior work has examined explainability in CPS and smart environments~\cite{abdul2018trends,miller2019explanation,sadeghi2021cases,sadeghi2024smartex}. A pioneering effort is the \emph{Intelligibility Toolkit}~\cite{lim2010toolkit}, which exposes APIs to generate multiple explanation types across several model families; however, as a developer toolkit (not a simulator) it lacks standardized user-study instrumentation.

%Platforms for smart-home simulation and evaluation also exist~\cite{houze2022generic,cook2009ambient}, though many trade scalability for realism. \emph{iCasa} is an OSGi/iPOJO environment that combines an Eclipse IDE, a smart-home simulator, and a scripting DSL for developing context-aware services~\cite{selfstar-website}. It targets platform development and autonomic context management rather and remains code-intensive. In contrast, V-SHiNE centers on smart-home simulation with explanations for user studies and emphasizes low/no-code authoring.

% ----------------------------
\section{Conclusion}
%V-SHiNE is currently optimized for desktop browser use and short-term studies. %While it simulates realistic environments, it cannot fully capture long-term, in-situ contextual nuances of real homes. 
%Despite these limitations, 
Our work demonstrates the feasibility and value of a lightweight, browser-based smart home simulation framework for evaluating explainability in CPS. The demonstration showcases how researchers can configure environments, inject behaviors, test explanation strategies, and collect comprehensive feedback. %Validation in two contexts highlights both feasibility and generalizability. 
By lowering the barrier for realistic, reproducible, user-centered evaluation, V-SHiNE provides a useful platform for advancing explainable intelligent systems and supporting future research in this area. Future work will extend support for multi-user interactions, richer device models, and long-term deployments. % as well as integrations with additional explanation engines. 
\section*{Tool Availability}
V-SHiNE is maintained on a public GitHub repository\\({\footnotesize\url{https://github.com/ExmartLab/SHiNE-Framework}}) and is licensed under the MIT license. 
It is easily deployable via Docker.
A comprehensive documentation \\({\footnotesize\url{https://exmartlab.github.io/SHiNE-Framework/}}) and a short demonstration video are available online ({\footnotesize\url{https://youtu.be/DmCuqt9DD18}}).

\bibliographystyle{IEEEtran}
\bibliography{references}

@inproceedings{houze2022generic,
  title={A generic and modular reference architecture for self-explainable smart homes},
  author={Houz{\'e}, Etienne and others},
  booktitle={Intl.\ Conference on Autonomic Computing and Self-Organizing Systems (ACSOS)},
  year={2022},
  organization={IEEE}
}

@inproceedings{sadeghi2021cases,
  title={Cases for explainable software systems: Characteristics and examples},
  author={Sadeghi, Mersedeh and Kl{\"o}s, Verena and Vogelsang, Andreas},
  booktitle={International Requirements Engineering Conference Workshops (REW)},
  pages={181--187},
  year={2021},
  organization={IEEE}
}

@article{baheti2011cyber,
  title={Cyber-physical systems},
  author={Baheti, Radhakisan and Gill, Helen},
  journal={The impact of control technology},
  volume={12},
  number={1},
  pages={161--166},
  year={2011}
}

@article{shin2021effects,
  title={The effects of explainability and causability on perception, trust, and acceptance: Implications for explainable AI},
  author={Shin, Donghee},
  journal={International journal of human-computer studies},
  volume={146},
  pages={102551},
  year={2021},
  publisher={Elsevier}
}

@article{doshi2017towards,
  author = {Finale Doshi-Velez and Been Kim},
  title = {Towards a rigorous science of interpretable machine learning},
  journal = {arXiv preprint arXiv:1702.08608},
  year = {2017}
}

@inproceedings{abdul2018trends,
  title={Trends and trajectories for explainable, accountable and intelligible systems: An hci research agenda},
  author={Abdul, Ashraf and others},
  booktitle={CHI conference on human factors in computing systems},
  year={2018}
}

@article{miller2019explanation,
  author = {Miller, T.},
  title = {Explanation in Artificial Intelligence: Insights from the Social Sciences},
  journal = {Artificial Intelligence},
  volume = {267},
  pages = {1--38},
  year = {2019},
  publisher={Elsevier}
}

@article{nunes2017systematic,
  title={A systematic review and taxonomy of explanations in decision support and recommender systems},
  author={Nunes, Ingrid and Jannach, Dietmar},
  journal={User Modeling and User-Adapted Interaction},
  volume={27},
  pages={393--444},
  year={2017},
  publisher={Springer}}

@misc{vermasan2014internet,
  title={Internet of things—from research and innovation to market development},
  author={Vermasan, Ovidiu and Friess, Peter},
  year={2014},
  publisher={River Publisher Series in Communication}
}

@inproceedings{baresi2018tdex,
  title={Tdex: A description model for heterogeneous smart devices and gui generation},
  author={Baresi, Luciano and Sadeghi, Mersedeh and Valla, Massimo},
  booktitle={International Conference on Internet of Things (IThings)},
  year={2018},
  organization={IEEE}
}

@misc{w3c-td,
  title         = {Web of Things (WoT) Thing Description 1.0},
  organization  = {World Wide Web Consortium (W3C)},
  note          = {W3C Recommendation},
  year          = {2020},
  month         = apr,
  howpublished  = {\url{https://www.w3.org/TR/wot-thing-description10/}},
  urldate       = {2025-09-27}
}

@article{jakobi2018evolving,
  author = {Jakobi, T. and Gunnarothers, S. and others},
  title = {Evolving needs in {IoT} control and accountability: A longitudinal study on smart home intelligibility},
  journal = {Proceedings of the ACM on Interactive, Mobile, Wearable and Ubiquitous Technology},
  year = {2018}
}

@inproceedings{carneiro2020explainable,
  title={Explainable intelligent environments},
  author={Carneiro, Davide and Silva, F{\'a}bio and Guimar{\~a}es, Miguel and Sousa, Daniel and Novais, Paulo},
  booktitle={International Symposium on Ambient Intelligence},
  pages={34--43},
  year={2020},
  organization={Springer}
}

@article{jha2022overview,
  title={An overview on the explainability of cyber-physical systems},
  author={Jha, Sanjiv},
  journal={Vol. 35 (2022): Proceedings of FLAIRS-35},
  year={2022},
  publisher={The Florida Artificial Intelligence Society}
}

@inproceedings{Reyd2024,
  title = {{CIRCE}: a Scalable Methodology for Causal Explanations in Cyber-Physical Systems},
  booktitle = {Intl.\ Conf.\ on Autonomic Comp. and Self-Organizing Systems (ACSOS)},
  author = {Reyd,  Samuel and Diaconescu,  Ada and Dessalles,  Jean-Louis},
  year = {2024},
}

\end{document}